\documentclass[preprint2]{aastex}
\usepackage{natbib}

\newcommand{\tyc}{TYC\,8380-1953-1}

\slugcomment{}

\shorttitle{The binary system TYC 8380-1953-1}
\shortauthors{L\'opez-Santiago et al.}

\begin{document}


\title{TYC 8380-1953-1: Discovery of an RS~CVn binary through the \textit{XMM-Newton} slew survey}


\author{J. L\'opez-Santiago}
\affil{Departamento de Astrof\'isica y Ciencias de la Atm\'osfera, 
        Facultad de Ciencias F\'isicas, 
        Universidad Complutense de Madrid,
        28040, Madrid, Spain}
\email{jls@astrax.fis.ucm.es}

\and

\author{B. Stelzer}
\affil{INAF - Osservatorio Astronomico di Palermo, 
        Piazza del Parlamento 1, 90134, Palermo, Italy}
\email{stelzer@astropa.inaf.it}

\and

\author{R. Saxton}
\affil{XMM-Newton Science Operations Centre European Space Astronomy Center (ESAC)
P.O. box (Apartado) 78
E-28691 Villanueva de la Ca\~nada, Madrid, Spain}
\email{Richard.Saxton@sciops.esa.int}




\begin{abstract}

In this paper we report the discovery of {the chromospherically active (RS CVn type) binary 
\tyc\ } through the \textit{XMM-Newton} slew survey and present results of our optical and X-ray 
follow-up.  
{With a flux limit of $6 \times 10^{-13}$ 
erg\,cm$^{-2}$\,s$^{-1}$ in the soft band ($0.2 - 2$ keV), the \textit{XMM-Newton} slew has a similar 
sensitivity to the \textit{ROSAT} All Sky Survey allowing interesting sources to be identified by their 
long-term variability.}
%
Two different types of stellar sources are detected in shallow X-ray surveys: young stars (both pre-main 
and main sequence stars) and chromospherically active binaries (BY Dra and RS CVn type systems). 
The discovery of stars in such surveys and the study of their nature through optical follow-ups is valuable 
to determine their spatial distribution and scale height in the Galaxy.
%
{Our analysis shows that \tyc\ is a double-lined spectroscopic binary with both components having similar spectral 
type (likely K0/2+K3/5) and luminosity. With a typical coronal temperature for an RS~CVn system 
($kT \sim 1.15$ keV) and an X-ray luminosity in the 0.3--10 keV energy band higher than $4 \times10^{31}$ erg\,s$^{-1}$, \tyc\ lies 
among the most X-ray luminous RS~CVn binaries.}

\end{abstract}


\keywords{Galaxy: stellar content -- X-rays: stars -- Stars: coronae --
Stars: chromospheres -- Stars: binaries: spectroscopic -- 
Stars: individual: TYC 8380-1953-1}

\section{Introduction}

The last decade has seen a wealth of wide-area surveys that have led to 
an explosive increase in our knowledge of the stellar census near the Sun. 
Despite these advances, our picture of our galactic neighborhood is far from
being complete. Three groups can be distinguished that represent the most 
frequent among the stars: flare stars, active binaries and isolated
young stars in the solar neighborhood. 
The census and space density of all three groups is unknown. 
\cite{Reid04.1} estimate that the $20$\,pc census of late-type 
dwarf stars, which include the dMe flare stars, 
is still incomplete to $\sim 20$\,\%, and obviously for larger 
distances there is a much higher fraction of missing stars. 
The most recent version of the {\it Catalog of chromospherically
active binary stars} \citep[CAB;][]{eke08} 
comprises 409 systems. Almost $50$\,\% of the stars in this post-{\it Hipparcos} 
version of the CAB are new identifications with respect to the previous 
catalog of \cite{Strassmeier93.1}. 
Finally, the systematic search of isolated young T Tauri stars, i.e. pre-main sequence stars located far
from present-day star-forming regions, started with the discovery of the 
prototypical object TW\,Hya \citep{Rucinsky83.1}. 
To date, the TW\,Hya association consists of a total of $30$ spectroscopically 
confirmed members, and the existence of $8$ other nearby young associations 
($d < 100$\,pc, age $<50$\,Myr) has been established \citep{Torres08.0}.
However, our knowledge of the stellar content of these associations is very incomplete 
due to their large extent of tens of square degrees on the sky resulting from the close
distance. 

A common feature of all three groups of stars is that they are characterized by 
strong magnetic activity and, therefore, they stand out in X-ray surveys. 
In the past, 
magnetically active stars have been discovered by the time-profile of their X-ray emission 
(outbursts lasting one to several hours), a thermal X-ray spectrum of $\sim 1$\,keV typical for stellar coronae 
and subsequent confirmation with optical spectroscopy 
\citep[e.g.][]{Hambaryan04.1, Kennea07.0}. 
Several types of astrophysical sources can cause variations of the
X-ray signal by factors of $100$ over short and/or long timescales. 
Therefore, for X-ray sources discovered during the \textit{ROSAT} All Sky Survey 
(RASS) or short snapshot observations, 
deep pointed X-ray follow-up observations provide constraints that are 
useful to discriminate between various candidate classes on basis of the luminosity, 
variability behavior and X-ray spectroscopy. 
Soft sources, e.g., must be associated with relatively low absorption and are 
likely of galactic origin.
Even if the stellar nature can be established through the X-ray properties, 
optical follow-up spectroscopy is needed for deriving, amongst
others, the spectral type, luminosity class, rotational and multiplicity state.  

We have followed this approach, and present here the discovery of an active
binary system from multi-epoch X-ray observations and subsequent optical
follow-up spectroscopy. 
Our target was recently identified as an X-ray bright object in the \textit{XMM-Newton} slew survey (XMMSL). 
Across the history of X-ray astronomy slew surveys have proven to be useful complements to
all-sky surveys and deep pointings.
For the first XMMSL catalog \citep{Saxton08.0} 
more than $200$ satellite slews have been analysed, covering $\sim 14$\,\% of the sky and 
yielding $2692$ X-ray sources down to a soft band ($0.2-2$\,keV) flux limit of
$6 \times 10^{-13}$ erg\,cm$^{-2}$\,s$^{-1}$. 
This is within a factor two of the sensitivity of the RASS. 
One of the most important contributions of slew surveys is the discovery of variable sources.
Various types of astrophysical sources were identified after optical follow-up of XMMSL transients,
e.g. supermassive black holes \citep{Esquej07.0} and novae \citep{Read08.0}.
We show here that the XMMSL can also be used to identify stars. 

In this work, we determine the nature of TYC 8380-1953-1, 
identified as a transient X-ray source in a recent \textit{XMM-Newton} slew. 
We have carried out a pointed \textit{XMM-Newton} observation of \tyc\ with the aim 
to detect its quiescent, i.e. non-flaring, X-ray emission. 
With two high-resolution optical spectra obtained with FEROS at the 2.2m ESO 
telescope on La Silla 
we intended to (1) infer its age by measuring the Lithium abundance ($6708$\,\AA),
(2) determine its rotation rate ($v \sin{i}$) as additional evidence for age/activity,
(3) and establish or refute its membership in any of the known young stellar associations
 in the solar neighborhood by studying the kinematics or find evidence for binarity
 from radial velocity (RV) variations. 



\section{Observations}\label{sect:obs}

\subsection{\textit{XMM-Newton} slew and previous knowledge on the source}\label{subsect:obs_slew}

On Sep 24, 2011 bright X-ray emission associated with TYC 8380-1953-1 
was detected during an \textit{XMM-Newton} slew (slew-ID. 9215900005; source-ID. 
XMMSL1\,J192252.3-483210). 
TYC 8380-1953-1 is a fairly unknown star with $B-V = 0.8$\,mag ($V=10.6\,$mag), 
corresponding to early-K spectral type for a dwarf. Making use of stellar evolutionary models, 
we estimated a photometric distance of $\sim 100$\,pc for an assumed age between 
$100$\,Myr and $1$\,Gyr. The same color for a giant star would correspond to an approximate spectral type 
G3. In such case, the star would be more distant \citep[$d \sim 900$ pc, assuming $M_\mathrm{V} = 0.9$ mag; see][]{landolt}. 
The only additional information available on this object are the coordinates
and proper motion from Tycho. The sky position of TYC 8380-1953-1 makes it a candidate
of the AB\,Dor association. 
The AB\,Dor group is one of the oldest young
nearby associations ($\sim 30-50$\,Myr; L\'opez-Santiago et al. 2006) 
and hosts several suspected RS\,CVn or BY\,Dra variables 
\citep[see summary in][]{Torres08.0}. 
These kind of variable stars are characterized by strong magnetic activity and fast rotation
related to synchronization with the orbital period in binary systems. 
Nothing was known about the multiplicity of TYC 8380-1953-1 prior to our campaign. 

{\tyc\ was detected with $17$ counts during an \textit{XMM-Newton} slew. 
All the slew observations are made with the EPIC/pn using the medium filter.
This slew source caught our attention because of its non-detection in
the RASS. 
We have extracted a spectrum from a circle of radius $30\arcsec$ centered on the source with the
background being extracted from an annulus of inner radius $60\arcsec$ and outer radius $90\arcsec$ 
around the source. The detector matrices were calculated taking into account
the transit of \tyc\ across the detector using the method described in \citet{Read08.0}.
%
A $0.3-2.0$\,keV count rate of $1.09 \pm 0.29$\,cts/s is obtained. Spectral fitting
is prohibited given the poor statistics.} However, we compare the slew 
spectrum to the best-fit model obtained from our pointed follow-up \textit{XMM-Newton}
observation (described in detail in Sect.~\ref{subsect:obs_point}). 
For this exercise, all spectral parameters
(temperature $kT$, column density $N_{\rm H}$ and abundance $Z$) were held fixed
at the values derived from the \textit{XMM-Newton} pointing and the normalization was
set to $N_{\rm slew} = N_{\rm poin} * CR_{\rm slew}/CR_{\rm poin}$ where 
$CR$ are the observed $0.3-2.0$\,keV count rates for the slew and the pointing,
respectively and $N$ the normalizations of the spectral model. The result is
displayed in Fig.~\ref{xray_fit_slew} and shows that the spectral shape during the \textit{XMM-Newton}
slew is compatible with that derived from the pointed observation. However,
during the slew the count rate -- and consequently the flux -- was about 
a factor three larger. 

The $2\,\sigma$ upper limit {from} the RASS {for} \tyc\ is $0.039$\,cts/s, 
which is equivalent to $0.31$\,cts/s in the energy band 0.3--2.0 keV for the 
\textit{XMM-Newton} EPIC/pn with thin filter
and $0.32$\,cts/s in \textit{XMM-Newton} EPIC/pn with medium filter, assuming the
spectral model from Sect.~\ref{subsect:obs_point}.  
This is on the order of the count rate measured during the \textit{XMM-Newton} pointing,
i.e. this upper limit does not provide any information on source variability. 


\subsection{\textit{XMM-Newton} pointed observation}\label{subsect:obs_point}

{We have been granted an \textit{XMM-Newton} target-of-opportunity observation of 
\tyc.} This observation was performed in October 13$^\mathrm{th}$
2011, during revolution 2169 (obs. id. 0679380601). The European Photon Imaging 
Camera (EPIC), operated in full-frame mode, was used as primary instrument. The 
exposure {times} for the \textsc{pn} and \textsc{mos} detectors {were} 5 and 6.6 ks, respectively. 
The X-ray background level remained constant at a low rate during the entire observing period. 
{This pointed \textit{XMM-Newton} observation was designed to be {$400$ times more
sensitive} than the \textit{XMM-Newton} slew data.} 

Data reduction was performed with the latest version of the XMM-\textit{Newton} specific 
software for data reduction and analysis ({Science} Analysis System, SAS, version 11.0). 
Filtered event lists for EPIC \textsc{mos} and \textsc{pn} were produced following the SAS 
standards. We do not use data from the Reflection Grating Spectrometer (RGS) 
because the source has {not enough counts}Ê to perform an analysis. The event
lists contain information on the arrival time of each photon detected during the observation, 
its energy and position on the detector.  
%
We used \textsc{evselect} to obtain the X-ray spectrum of the source identified with the 
star TYC 8380-1953-1. We chose a circular extraction region of radius $27\arcsec$
encircling 80\% of photons in the $0.3-10.0$\,{ keV} EPIC energy band.
Background was extracted from a local source-free region using the same 
extraction region radius than for the source.
The light curve of TYC 8380-1953-1 was extracted from the X-ray event list using the 
same extraction region than for the {spectrum}, with an IDL code 
specifically created by us. 
The analysis of the X-ray spectrum and light curve of our target is presented in 
Section~\ref{analysis_x}.

\subsection{Optical spectroscopy}
\label{optical_obs}

{Optical} spectroscopic observations of TYC 8380-1953-1 were carried out on 
9$^{th}$ and 10$^{th}$ November 2011. 
We used the high-resolution spectrograph FEROS, mounted on the 2.2m telescope of {the}
La Silla Observatory of the European Southern Observatory. The chosen configuration provides
$R \sim 50000$ at 6562 \AA, measured from the FWHM of comparison-arc emission lines. FEROS 
covers the entire optical spectral range from 3650 to 8900 \AA.

One spectrum of the star was taken during each 
{ of the two nights}, together with { a spectrum of} the reference star $\beta$ Ceti, 
a spectral type and radial velocity standard K0 giant.
The exposure time for the target was 634 s in both observations, giving a
signal-to-noise ratio $S/N \sim 30$ in H$\alpha$, Na~\textsc{i} doublet and H$\beta$ lines and 
$S/N \sim 10$ in Ca~\textsc{ii} H \& K. The maximum {$S/N$ ratio in those spectra is 
$\sim 55$}, that was reached at red wavelengths ($\lambda > 7000$ \AA). 
The $S/N$ for the reference star ranges from 100 to 200, from blue to red wavelengths. 

The { data} reduction process was performed using the standard procedures in the 
IRAF\footnote{IRAF is distributed by the National Optical Observatory, which is operated 
by the Association of Universities for Research in Astronomy, Inc., under contract with the 
National Science Foundation.} package (bias subtraction, extraction of scattered light produced 
by optical system, division by a normalized flat-field and wavelength calibration). After reduction, 
each spectrum was normalized to its continuum, order by order, by fitting a polynomial function.
Despite the good quality of the flat-field spectra, the fringing pattern {could not be completely 
removed} from the stellar spectra. This problem affected only long wavelengths of the 
star's spectrum, where the Ca~\textsc{ii} IR-triplet is located. Therefore, no analysis  
of those spectral lines was done. 

\section{Data analysis}
\label{analysis}

\subsection{X-ray spectrum and light curve}
\label{analysis_x}

For the analysis of the X-ray spectrum of TYC 8380-1853-1, we used the C-code based software 
XSPEC \citep[v.12; ][]{arn04} with a plasma model calculated using the Astrophysical Plasma Emission 
Database (APED), that is included in  AtomDB\footnote{http://www.atomdb.org/}.  
The reference spectrum generated by XSPEC is that of a collisionally-ionized diffuse gas. To account
for the interstellar and/or circumstellar absorption in the line of sight of the source, XSPEC permits
to use an absorption (multiplicative) model in combination with the plasma model. For our analysis, 
we used \textit{wabs}, a photo-electric absorption law determined with the Wisconsin \citep{mor83} 
cross-sections.

As observed in the Sun, the corona of any star shows an X-ray spectrum that is fully characterized 
by its emission measure distribution. 
At the energy resolution of the EPIC, the spectrum of a coronal source is well-represented by one 
or two thermal components \citep{bri03,lop07}. Multi-temperature models with three or more thermal components 
are needed only when the statistics very high \citep[e.g.][]{rob05,lop10c}. 

For our target, \tyc, we collected $\sim 3000$ counts 
in total summing up photons of the PN and MOS
cameras in the energy range 0.3--10.0 keV,  
The spectrum is shown together with the best one- and two-temperature fits
in Fig.~\ref{xray_fit_spec} and Table~\ref{xray_fit_tab}
summarizes the {best} fit parameters.
The reduced $\chi^2$ (column 7 of Table~\ref{xray_fit_tab}) shows that the 
one-temperature model provides a good description of the spectrum. 
Adding a second thermal component to the model
does not give further information, except for a better fit to the hard X-ray 
tail of the spectrum. 
However, the additional hot temperature is not well constrained 
and this component contributes very little to the spectrum in the EPIC 
energy range {({the} emission measure ratio between both thermal components 
is $EM_\mathrm{1}/EM_\mathrm{2} \sim 7$; see Table~\ref{xray_fit_tab}).} 
The value of the column density measured from the X-ray spectrum
is comparable with the Galactic column density at 1 Kpc in the line of sight out of the 
Galactic disk \citep{kal05}. 
This result suggests that the star is not nearby 
(see also Sections~\ref{spt_vr_vsini} and \ref{conclusions}). 

In Fig~\ref{xray_lc}, we plot the EPIC/\textsc{pn} lightcurve of \tyc\ in $0.3-10$\,keV. 
Bins in the figure are four minutes long. The lightcurve is background 
corrected. A Kolmogorov-Smirnov test gives very high probability for the 
star's count-rate to be constant { throughout the} observation ($P > 95 \%$).
The observation is {by far} not long enough to detect rotational modulation or eventual variations during 
orbital phase; {see Sect.~\ref{conclusions} for an estimate of the period}.

\subsection{High-resolution optical spectrum}

A simple visual comparison of the {spectra} of \tyc\ observed during 
both nights shows that the star is a double-lined spectroscopic binary system 
(SB2; see Fig.~\ref{ap11}). 
During the first observing night, the spectral lines of the two stars in the 
system were blended. During the second night, both spectral components were 
clearly separated. 
The strength of iron and calcium lines of the secondary star is slightly lower 
than that of the primary but the ratio between lines in both stars is very similar, 
what suggests that both stars have similar spectral types. 
{ In the remainder of this section we determine spectral type, RV and rotation
rates and examine the chromospheric activity of both components in the binary.}
{Tables~\ref{jstarmod} and \ref{param}} summarize the results obtained from 
the optical spectral analysis.

\subsubsection{Spectral types, radial velocities and rotation}
\label{spt_vr_vsini}

To determine the spectral types of both stars in the { binary} system, 
we followed \citet{lop10b}. 
We first studied different spectral type and luminosity class indicators, 
including spectral line ratios and equivalent widths. In particular, we used the 
lines Ti~\textsc{i} $\lambda5866$ \AA, the Mg~\textsc{i} triplet at 
$\lambda5167$, 5172 and 5183 \AA, Fe~\textsc{i} $\lambda6430$ \AA, 
Fe~\textsc{ii} $\lambda6432$ and 6457 \AA, Ca~\textsc{i} $\lambda6439$, 6449 and 
6456 \AA, Co~\textsc{i} $\lambda6455$ \AA and V~\textsc{i} $\lambda6452$ \AA, as 
described in \citet{str90}, \citet{mon07} and \citet{gra09}. From the different 
line ratios, we conclude that both stars in the binary system are of 
early-K spectral type. However, the large broadening of the line profiles 
prevents the precise determination of the spectral sub-types. We tentatively 
classify the system as K0/2+K3/5.   

The Na~\textsc{i} doublet at 5890 \AA\ is a good indicator of gravity
for some spectral types \citep[e.g.][]{mon99}. In particular, the wings of these 
lines become narrower in evolved K and M stars (giants or subgiants). Also good 
indicators of gravity are strong lines like the Ca~\textsc{i} lines at 4227 and 6162 \AA\ 
\citep{gra09}. In our spectra those lines are blended with other spectral
lines due to rotational broadening. The situation is even worse because \tyc\ is a 
SB2 system and the lines of both components are present in the spectra. Other 
typical indicators of gravity such as the Balmer series and the Ca~\textsc{ii} H \& K
lines are not useful for active stars since their profiles are affected by the presence of 
chromospheric emission. The profile of the Na~\textsc{i} doublet lines indicates 
that both components of \tyc\ are evolved. This hypothesis is confirmed 
during the spectral subtraction process (see below).
The spectrum of \tyc\ at the position of the Na~\textsc{i} doublet during the second night 
is shown in Fig.~\ref{nai}. During this observation, it showed the lines of 
both stars in this binary system and a narrow absorption line produced by the interstellar
medium present in between the star and the Earth. This is demonstrated in Fig.~\ref{nai}
bottom, where we show the spectrum that remains if the photospheric spectrum of the two 
stars in the binary are subtracted (see bellow for a description of the subtraction technique).
We have measured the equivalent width of these {interstellar} Na~\textsc{i} lines, 
{and obtained} $EW = 0.148 \pm 0.006$ \AA. Using the relations of \citet{mun97}, 
$E \mathrm{(B-V)} \approx 0.05 \pm 0.01$ mag ($A_\mathrm{V} = 0.16 \pm 0.01$ mag), 
corresponding to a column density $N_\mathrm{H} = 3.6 \pm 0.1 \times 10^{20}$ cm$^{-2}$
\citep[see][]{gor75}. 
This value is { marginally compatible with} that obtained from the X-ray { spectral} fitting 
($N_\mathrm{H} \approx 7 \pm 3 \times 10^{20}$ cm$^{-2}$; see Table~\ref{xray_fit_tab}).

To complete our analysis of the optical {spectra} of TYC 8380-1853-1, 
we performed spectral subtraction of a spectral type standard star 
on our target using 
\textsc{jstarmod}, a perl-plus-fortran code based software that permits to combine the 
spectra of different objects and perform subtraction on the spectrum of another object
\citep[the interested reader will find a detailed discussion on the spectral subtraction 
technique and the use of \textsc{jstarmod} in][Sections 3.4 and 3.5]{lop10b}.
The spectral type standard K0 giant $\beta$ Ceti was used as the reference
star. $\beta$ Ceti was observed together with our target during each night. \textsc{jstamod} permits 
to use more than one reference star (with different weights) to reproduce the spectrum of 
binary systems. We used $\beta$ Ceti as the reference star for both components of 
TYC 8380-1853-1, since they have very similar spectral types.
The fit was performed separately for each night. The results of the fits are shown in 
Table~\ref{jstarmod}. The rotational velocities of { both} components 
for the first night (where their spectral lines where blended) were fixed to the values obtained
for the second night. The radial velocities given in Table~\ref{jstarmod} are relative to the 
standard star. 
{ Additional} observations { are} needed to determine precisely the 
binary orbit and stellar parameters, { in particular the} stellar masses. 
The rms values given in 
column 8 of Table~\ref{jstarmod} were measured in the subtracted spectrum and they 
account for the errors in the fitting as well as for possible differences in the spectral 
type of the two stellar components of TYC 8380-1953-1 and the reference star
$\beta$ Ceti \citep[see][for details]{lop03}.

\subsubsection{Chromospheric activity}

The spectrum of TYC 8380-1953-1 shows several chromospheric lines in emission. The 
H$\alpha$ line shows an inverse P Cygni profile caused by the superposition of the absorption 
line of the primary and the emission line of the secondary (see Fig.~\ref{ha}). In fact, the 
primary presents a filled-in H$\alpha$ line instead of a { pure} absorption line. This is easily 
seen when the spectrum is compared with that of the reference star  (Fig.~\ref{ha_sub}). 
We used $\beta$ Ceti as reference for the subtraction of the photosphere of both 
components of TYC 8380-1953-1. Radial and rotational velocities were fixed to the values 
{ determined in Section~\ref{spt_vr_vsini}}. 
We then fitted the subtracted H$\alpha$ spectrum of each night { with} two Gaussian 
functions using the IRAF task \textsc{splot}. The results are shown in Fig.~\ref{ha_gauss} and 
Table~\ref{param}. The values given in Table~\ref{param} 
are corrected for the different contribution of each star to the continuum. 

The H$_\beta$ line also shows { some evidence for filling-in of} its core  
but the subtracted spectrum is { too} noisy to perform a { quantitative} analysis.
The Ca~\textsc{ii} H \& K lines { are} detected in emission, but the low 
S/N ratio in this specific spectral region in our observations prevented us 
{from} applying the spectral subtraction technique. 
In Fig.~\ref{cah} we compare the observed spectrum of \tyc 
and $\beta$ Ceti in the region of the Ca~\textsc{ii} H line. { No} other 
chromospheric lines { are} detected in emission, { nor are there any signs 
for filling-in of} photospheric lines of the Balmer series { other than H$\beta$}. 
However, the presence { of weak emission lines} in the spectrum of \tyc\ cannot be 
{ excluded} as the { S/N} ratio of our observation is not high 
enough to detect them.

\section{Discussion and conclusions}
\label{conclusions}

{Precise classifications of suspected galactic stars are important for a correct 
assessment of the space density of various types of stars, such as flare stars,
young nearby stars and active binaries.
Our X-ray and optical follow-up observations of the \textit{XMM-Newton} slew transient
XMMSL1\,J192252.3-483210, identified with the Tycho source \tyc, have allowed us 
to constrain the nature of this star. 
} 

First, we can exclude that \tyc\ is a flaring dwarf star (dMe) in the 
solar neighborhood because of the signatures of low gravity found in the
optical spectrum (narrow Na~\textsc{i} doublet).
Secondly, we can discard the hypothesis that \tyc\ is a member of a young stellar 
association or moving group { in the solar neighborhood} based on the following
arguments. 
Assuming a typical absolute magnitude  
for a giant K star ($M_\mathrm{V} = 0$) and that both components have very similar 
luminosity ($V = 10.6$ mag for the binary system), the distance to \tyc\ 
{ can be estimated to be} close to $2$\,kpc
($d \sim 1$~kpc for $M_\mathrm{V} = 2$, typical of less luminous giants and 
subgiants). Even in the most optimistic case (assuming { that} 
the stars are low luminous subgiants), the distance to \tyc\ would be 
approximately $600$\,pc. 
Moreover, 
{ both} the column density determined from the equivalent width of the 
narrow Na~\textsc{i} doublet components observed in the 
stellar spectrum { and that} obtained 
from the X-ray spectrum, indicate significant
interstellar extinction, i.e. the system is not nearby. 
Finally, another proof against this hypothesis is the { absence} of the lithium 
absorption line at 6708\,\AA. 

{ Our sequence of two optical spectra obtained in two consecutive
nights has shown strong RV variations indicating binarity. Since the 
stars in the system are evolved,  we conclude that
\tyc\ is an RS~CVn binary.} 
These kind of systems typically 
consist of an evolved chromospherically active cool star (spectral type G or later) and a hotter component, 
evolved or not and usually {inactive} \citep{fek86}. However,
several RS~CVn systems {consist of two evolved stars of similar spectral type}
\citep[see the chromospherically active binaries (CAB) catalog by][]{eke08}, 
{ as we observed for \tyc\ where}
both components have { early-K} spectral type and present chromospheric 
activity { evidenced by the presence of optical emission lines}. 
The stellar parameters of TYC 8380-1853-1 are compatible with other chromospherically
active binary systems such as DK Dra, { an SB2} of { spectral} type K1~III + K1~III
\citep{mon00}. {Similar to \tyc}, DK Dra shows filled-in H$\alpha$ profiles and Ca~\textsc{ii} H \& K 
in emission for both components. A recent list of {typical} parameters of chromospherically active binaries
is compiled in the CAB catalog. 

The X-ray properties of \tyc\ are also consistent with this classification
as an active binary. At a distance between $600$\,pc and $2$\,kpc, the observed X-ray flux during
the \textit{XMM-Newton} {pointing} corresponds to an X-ray luminosity of $4 \times 10^{31}$ and 
$5 \times 10^{32}$ erg\,s$^{-1}$, respectively, at the upper end of the range observed for RS~CVn systems 
in the CAB. 
It is definitely higher than that of nearby K-type field dwarf stars \citep[$10^{27.5}$\,erg/s;][]{Schmitt04.1}. 
{It is also at the upper end of the range observed for the RS~CVn systems  that have values in
the CAB. 
This can be seen in Fig.~\ref{lx_vsini}
where we show \tyc\ (vertical bar) together with various sub-samples of RS~CVn
systems listed in the CAB.} This result is in favor of the lower 
{of the two distances given above} for \tyc.
During the \textit{XMM-Newton} slew, the X-ray luminosity was a factor of three higher than during
the pointed observation. The star likely {underwent a flare during that time}. Although 
this hypothesis cannot be reliably demonstrated with the current data, the X-ray luminosity of 
\tyc\ during the \textit{XMM-Newton} slew is {higher than the flare peak luminosities reported} 
by \citet{pan12} for other RS CVn binary systems. 

The coronal temperature of TYC 8380-1953-1 (k$T \sim 1.15$ keV, $T \sim 12$ MK) 
is typical of both RS~CVn and BY~Dra binary systems \citep{dem93, dem93a}, although
similar coronal temperatures are also observed in {single} late-type dwarf stars 
\citep[e.g.][]{fav03, g04}. 
{ Generally, the X-ray luminosity of late-type stars is linked with the rotation
period in a so-called rotation-activity relation where fast rotation goes along
with strong X-ray emission 
\citep{Pallavicini81.1, Pizzolato03.1}. 
{There is no consensus in the literature on the question 
if RS~CVn binaries follow a
rotation-activity connection \cite[e.g.][]{Walter81.3, Fleming89.1}. Possibly this
is due to the incompleteness and biases of all RS~CVn samples examined so
far. Fig.~\ref{lx_vsini} demonstrates this on the basis of the 
most recent compilation of RS~CVns in the CAB catalog 
The total CAB sample is biased towards stars with high X-ray luminosity,
which are absent in the known $50$-pc sample. The restricted
sample within $50$\,pc, presents some evidence for a rotation-activity
connection (no low-activity stars have fast rotation). Note, that possibly not 
even the $50$\,pc is complete, and that systems with a range of spectral types
are included in the CAB. In fact, this catalog 
contains only three \tyc\ analogs, i.e. RS~CVn binaries composed of two
K-type stars. These are highlighted with star symbols in Fig.~\ref{lx_vsini}.
We conclude that \tyc\ is the most active of the known systems in a rare group 
of active binaries.} 

Finally, 
{the} chromospheric emission level observed for both components of \tyc\ is similar to 
that obtained for chromospherically active binaries, {whether} evolved or not \citep[e.g.][]{mon00,gal07}. 
{ We can speculate on the origin of the} large difference in the equivalent 
width of H$\alpha$ ({see Table~\ref{param}}) of the secondary star { for the 
two nights. One possibility is that its chromospheric emission level 
may have changed because active regions have moved into the 
line-of-sight as a result of rotation. (Recall that the $25$\,h interval
between the two spectra likely represents a significant fraction of the stellar
rotation cycle). Another possibility is that} during the first night, the primary 
star was eclipsing the secondary. { Last but not least,} 
it is also { possible} that both stars {have} 
chromospheric activity in that part of the stellar disk facing the other star.
{ However, the periodic flaring in loop systems in between two close binary 
stars predicted by MHD simulations \citep{Gao08.1} have not been confirmed
by observations \citep{Stelzer02.1}. } 

{The discovery of RS~CVn type binaries, such as \tyc, is important 
because} the Galactic scale height of these objects is poorly known. In several 
works \citep[e.g.][]{fav03, mic07, lop07} 
it was noticed that shallow X-ray surveys show an excess of `yellow' stars when 
compared to Galactic models. In optical follow-ups of the stellar X-ray sources in those 
surveys, several authors have demonstrated that part of that excess is produced by 
the presence of evolved binaries, that are not well accounted for by the present Galactic 
models \citep[][L\'opez-Santiago et al. 2012 in prep.]{sci95, aff08} 
We have here {identified} one such binary system.



\acknowledgments

JL-S acknowledges financial support by the Spanish Government under grants 
AYA2011-30147-C03-02 and AYA2011-29754-C03-03. JL-S also 
thanks project AstroMadrid (S2009/ESP-1496) for partial support. 
BS thanks S. Sciortino for discussions on the activity on RS CVn binaries. 
BS acknowledges financial support from ASI/INAF grant n. I/035/10/0-10/308.
The authors would like to acknowledge the XMM-Newton Science Operations Team 
for the ToO observation of this source. We also acknowledge the ESO DDT Committee 
for the completion of the DDT Program ID 288.D-5006. 
We would like to thank the referee for useful comments and suggestions. 




{\it Facilities:} \facility{ESO/La Silla 2.2m (FEROS)}, \facility{XMM-Newton (EPIC)}.

\bibliography{yaReferences}
\bibliographystyle{pasp}


\clearpage


\begin{table*}
\caption{Results of the spectral fit of a 1T- and 2T-models to the observed X-ray spectrum of
TYC 8380-1953-1. Errors are at 90\% confidence level ($\Delta\chi^2 = 2.7$).}
\scriptsize
\begin{center}
\begin{tabular}{ccccccccccccccccc}
\hline\hline
\noalign{\smallskip}
Numer of thermal &  $N_\mathrm{H}$ &  $kT_1$ & $kT_2$ & $EM_1/EM_2$ & Z & $\chi^2_\textrm{red}$ & 
    $f_\mathrm{X}^\mathrm{observed}$ & $f_\mathrm{X}^\mathrm{unabsorbed}$ \\
\noalign{\smallskip}
  components &$(\times10^{22}$ cm$^{-2})$ & (keV) & (keV) &  & (Z$_{\odot}$) & (d.o.f) & (erg cm$^{-2}$ s$^{-1}$) & (erg cm$^{-2}$ s$^{-1}$) \\
\noalign{\smallskip}
\hline 
\noalign{\smallskip}
1 & 0.07$_{-0.01}^{+0.03}$ & 1.15$_{-0.14}^{+0.07}$ & \nodata & \nodata & 0.10$_{-0.05}^{+0.03}$ & 1.06 (176) & $7.0 \times10^{-13}$ & $9.2 \times10^{-13}$  \\
%
\noalign{\smallskip}
2 & 0.07$_{-0.01}^{+0.02}$ & 0.99$_{-0.06}^{+0.05}$ & 17.2$_{-13.4}^{+17.2}$ & 6.9 & 0.11$_{-0.04}^{+0.05}$ & 0.86 (174) & $8.4 \times10^{-13}$ & $1.0 \times10^{-12}$  \\
\noalign{\smallskip}
\hline
\end{tabular}
\end{center}
\label{xray_fit_tab}
\end{table*}

\begin{table*}
\caption{Results of the cross-correlation of TYC 8380-1953-1 with the K0~III 
standard star $\beta$ Ceti.}
\scriptsize
\begin{center}
\begin{tabular}{ccccccccccccccccc}
\hline\hline
\noalign{\smallskip}
Night & \multicolumn{2}{c}{Radial velocity} & & \multicolumn{2}{c}{Rotational velocity} & & 
\multicolumn{2}{c}{weight} & rms \\
\noalign{\smallskip}
\cline{2-3}\cline{5-6}\cline{8-9}
\noalign{\smallskip}
   & primary & secondary & & primary & secondary & & primary & secondary & \\
\noalign{\smallskip}   
   & (km s$^{-1}$) & (km s$^{-1}$) & & (km s$^{-1}$) & (km s$^{-1}$) & 
   & (km s$^{-1}$) & (km s$^{-1}$) & \\
\noalign{\smallskip}
\hline
\noalign{\smallskip}
1  & 12.4 & ~36.7 & & 34.1 & 24.8 & & 0.6 & 0.4 & 0.02 \\
\noalign{\smallskip}
2  & 21.5 & -54.3 & & 34.1 & 24.8 & & 0.6 & 0.4 & 0.02 \\
\noalign{\smallskip}
\hline
\end{tabular}
\end{center}
\label{jstarmod}
\end{table*}

\begin{table*}
\caption{Summary of the stellar parameters of TYC 8380-1953-1 from the analysis of the 
high-resolution spectrum.}
\scriptsize
\begin{center}
\begin{tabular}{lcc}
\hline\hline
\noalign{\smallskip}
Parameter & \multicolumn{2}{c}{value} \\
\noalign{\smallskip}
\cline{2-3}
\noalign{\smallskip}
   & primary & secondary \\
\noalign{\smallskip}
\hline
\noalign{\smallskip}
Spectral type & K0/2 & K3/5 \\
\noalign{\smallskip}
Luminosity class & IV/III & IV/III \\
\noalign{\smallskip}
$v \sin i$ [km~s$^{-1}$] & 34 $\pm$ 3 & 25 $\pm$ 2 \\
\noalign{\smallskip}
$EW\mathrm{(H_\alpha)}$ [\AA] (Nov 9) & 0.92 $\pm$ 0.04 & 1.08 $\pm$ 0.08 \\
\noalign{\smallskip}
$EW\mathrm{(H_\alpha)}$ [\AA] (Nov 10) & 0.59 $\pm$ 0.03 & 3.30 $\pm$ 0.09 \\
\noalign{\smallskip}
\hline
\end{tabular}
\end{center}
\label{param}
\end{table*}


\begin{figure*}
\includegraphics[width=17cm]{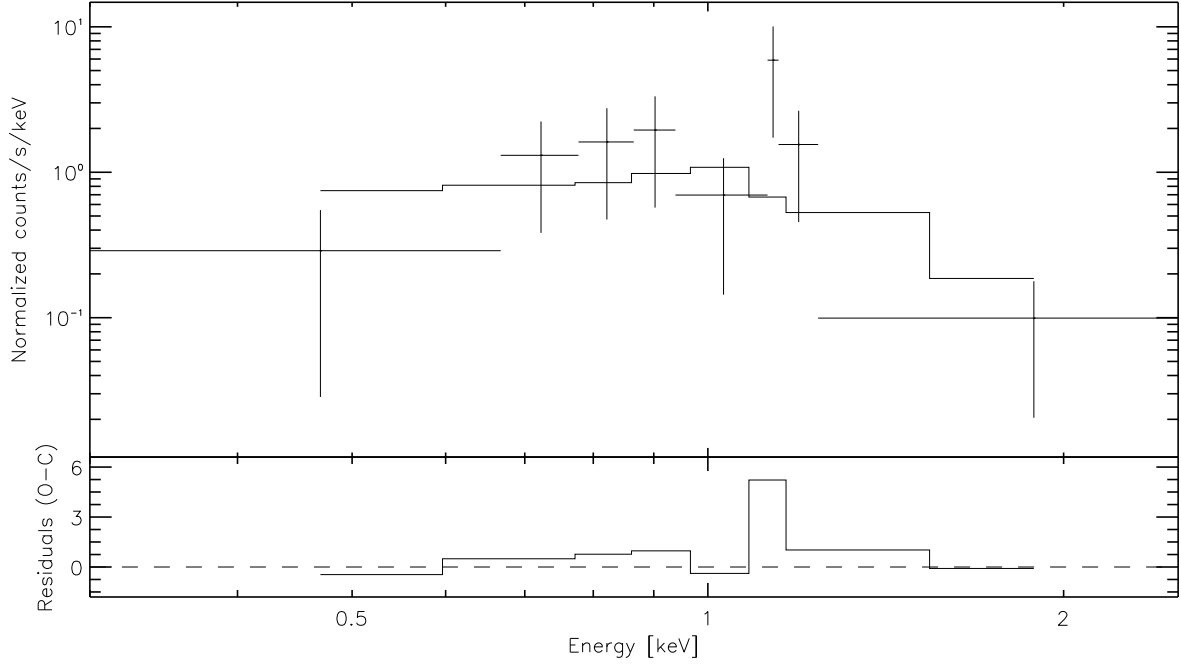}
\caption{X-ray spectrum of \tyc during XMM-\textit{Newton} slew compared to the 
best-fit model obtained from our pointed observation. The normalization is scaled
to $N_{\rm slew} = N_{\rm poin} * CR_{\rm slew}/CR_{\rm poin}$. The remaining 
parameters are fixed to those obtained from our pointed observation.  
\label{xray_fit_slew}}
\end{figure*}

\begin{figure*}
\includegraphics[width=17cm]{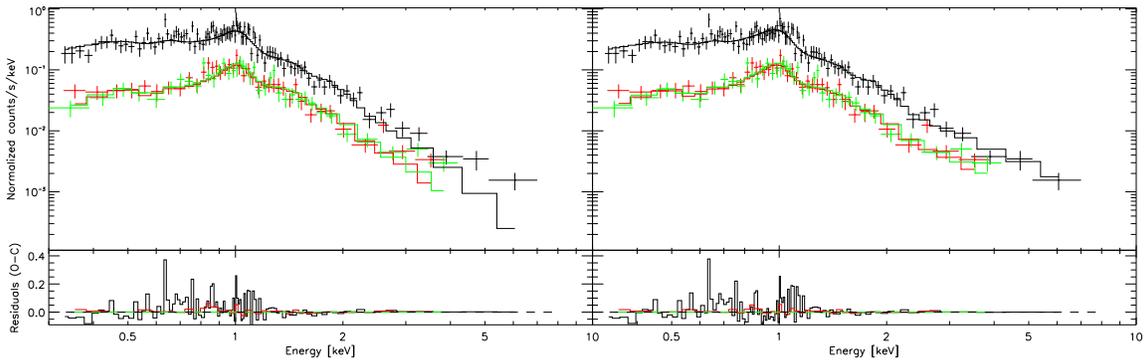}
\caption{EPIC PN { (black)} and MOS { (red and green)} spectra of 
\tyc\ in the energy range $0.3-10.0$\,keV. Continuous lines { represent} the 
model fitted { with} XSPEC using plasma models with one temperature (left panel) and two temperatures 
(right panel), respectively. 
\label{xray_fit_spec}}
\end{figure*}

\begin{figure*}
\includegraphics[width=17cm]{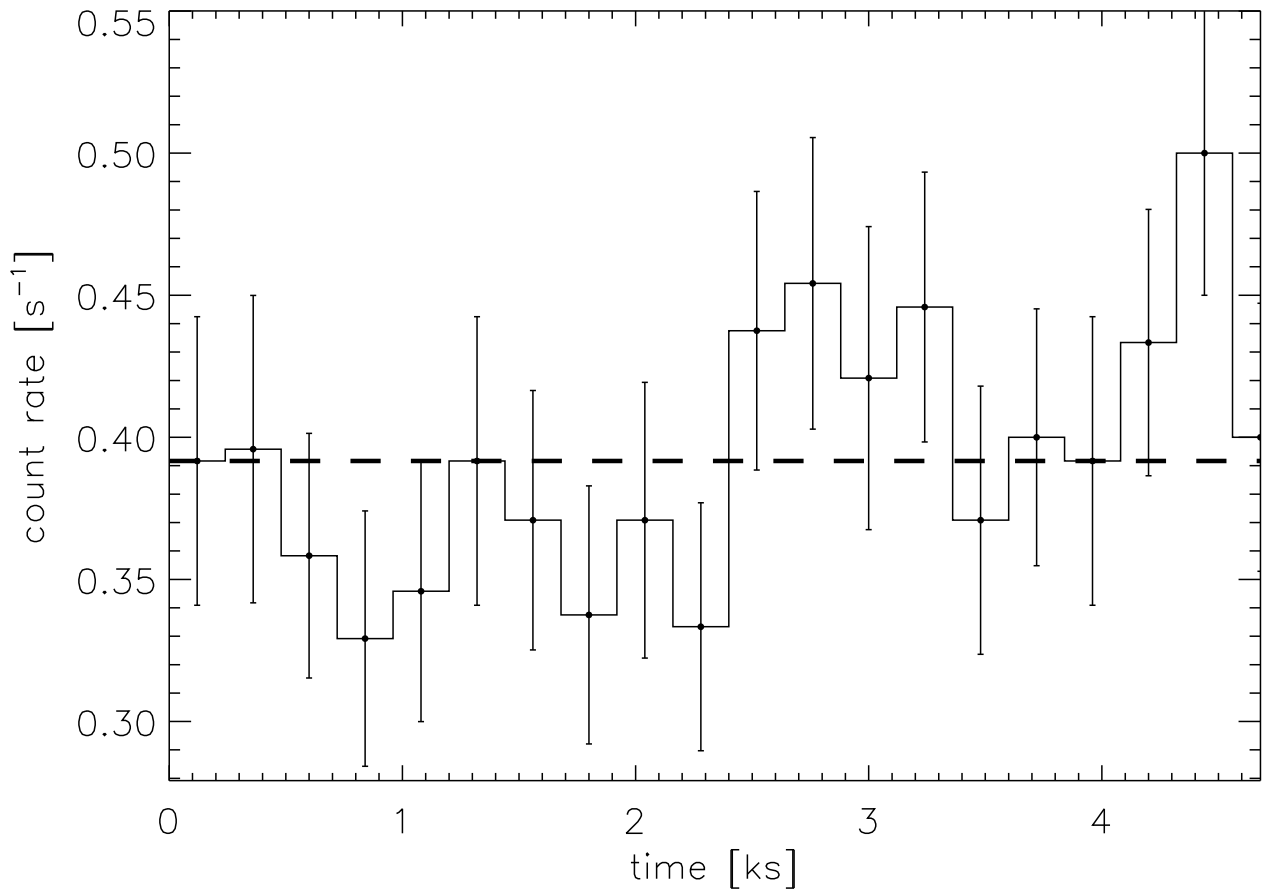}
\caption{X-ray lightcurve { of \tyc}\ in the $0.3-10$\,keV energy range 
{ observed with} EPIC PN. Time bins are 4 minutes long. The dashed line 
is the mean count-rate of the source. 
\label{xray_lc}}
\end{figure*}

\begin{figure*}
\includegraphics[width=17cm]{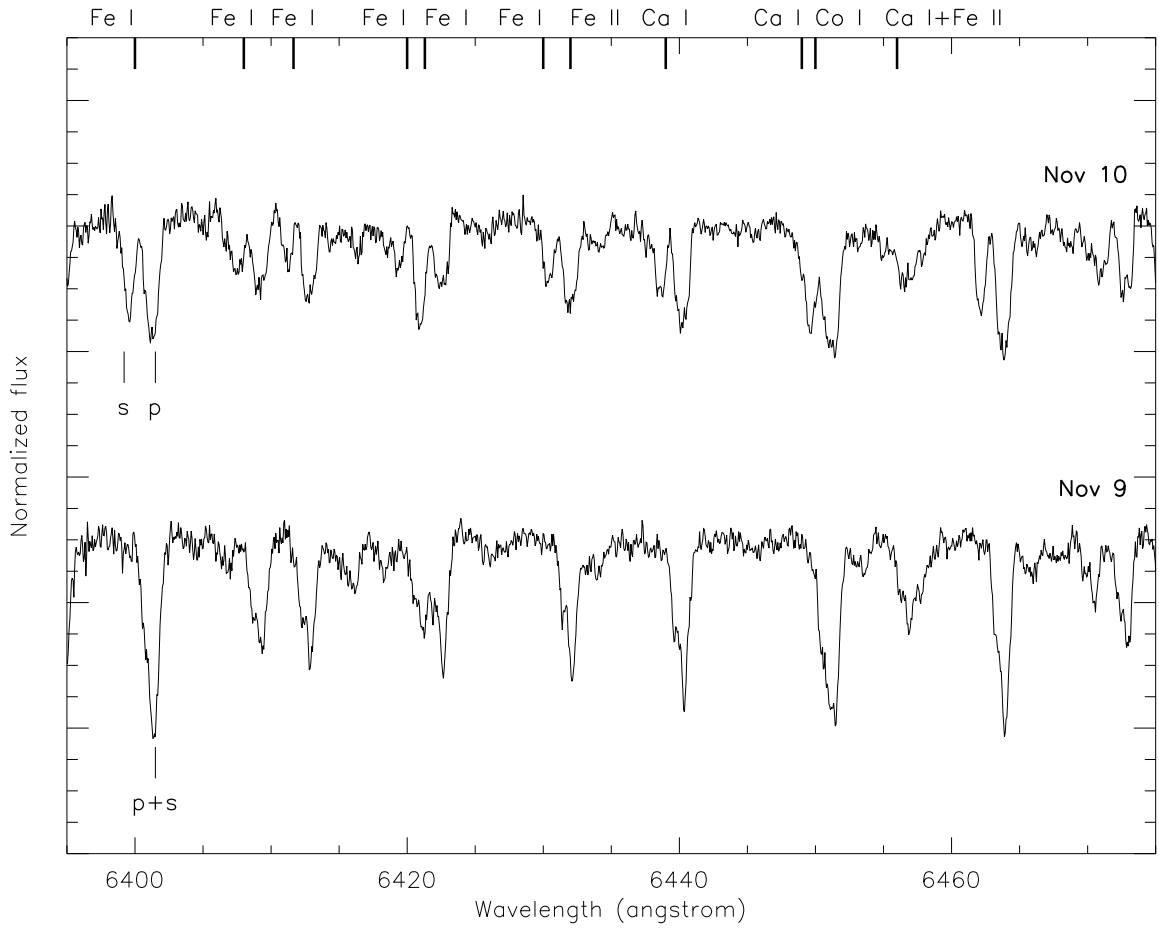}
\caption{Normalized spectrum of \tyc\ close to the H$\alpha$ line
{observed on Nov 9 and 10, 2011}. All the lines observed in 
{the displayed part of the} spectrum are from Ca~\textsc{i} and Fe~\textsc{i}.
We mark laboratory position for some representative lines.
\label{ap11}}
\end{figure*}

\begin{figure*}
\includegraphics[width=17cm]{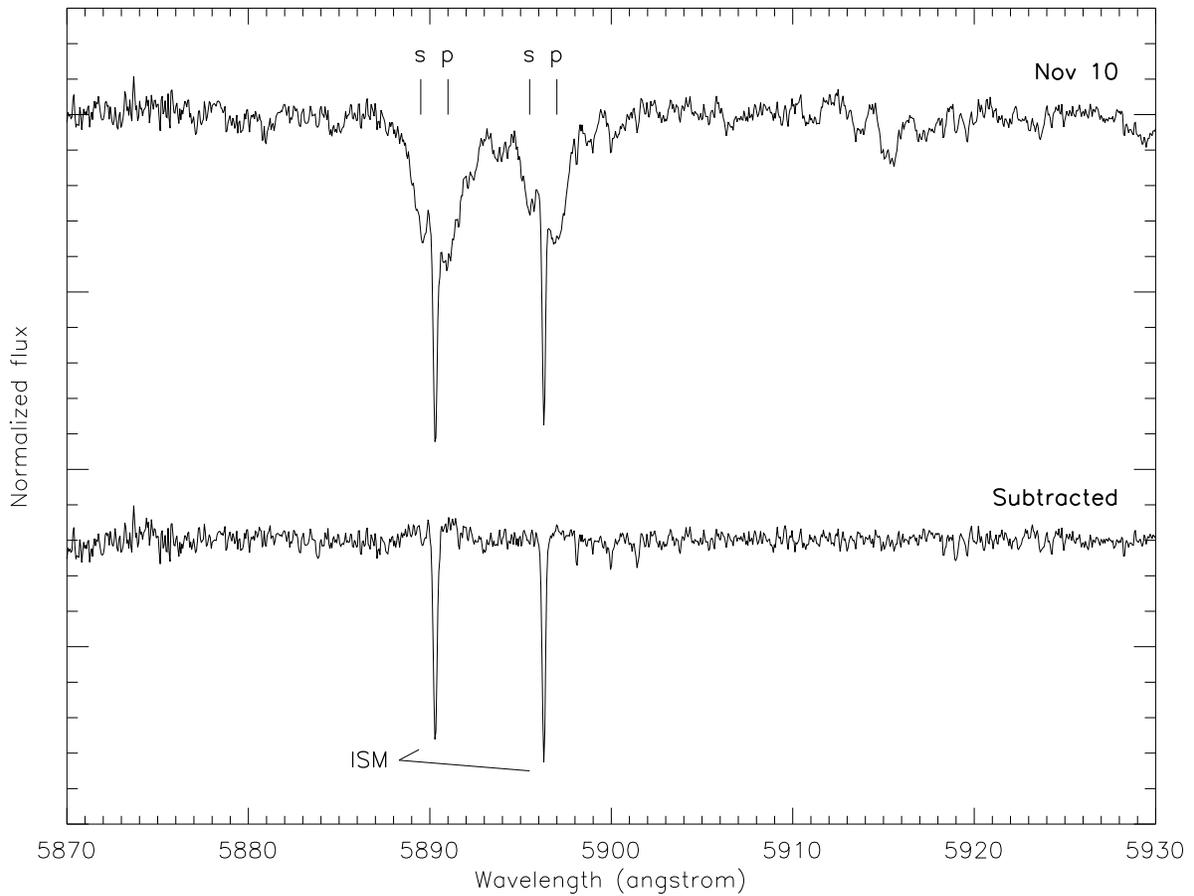} 
\caption{Normalized spectrum of \tyc\ during Nov 10, 2011 showing  
the Na~\textsc{i} doublet of both stars in the system and the 
narrow-absorption component associated with interstellar medium.
The spectrum at the bottom is the result of subtracting the synthetic 
photospheric spectrum of \tyc\ created with \textsc{jstarmod} from the 
observed spectrum of $\beta$ Ceti (see Section~\ref{spt_vr_vsini}
for details).
\label{nai}}
\end{figure*}

\begin{figure*}
\includegraphics[width=17cm]{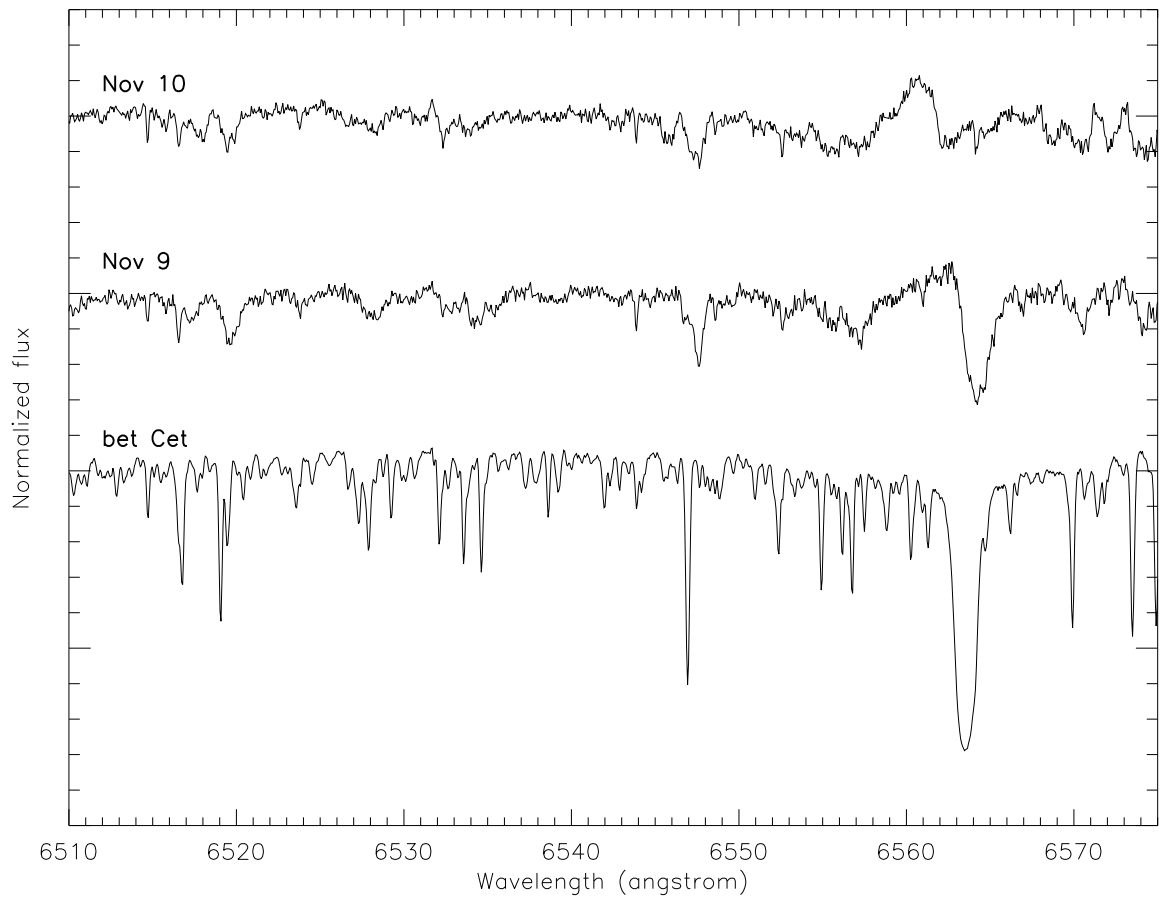} 
\caption{Normalized { spectra} of \tyc\ and $\beta$\,Ceti in the region of 
the H$\alpha$ line during Nov 9 and 10, 2011.}
\label{ha}
\end{figure*}

\begin{figure*}
\includegraphics[width=17cm]{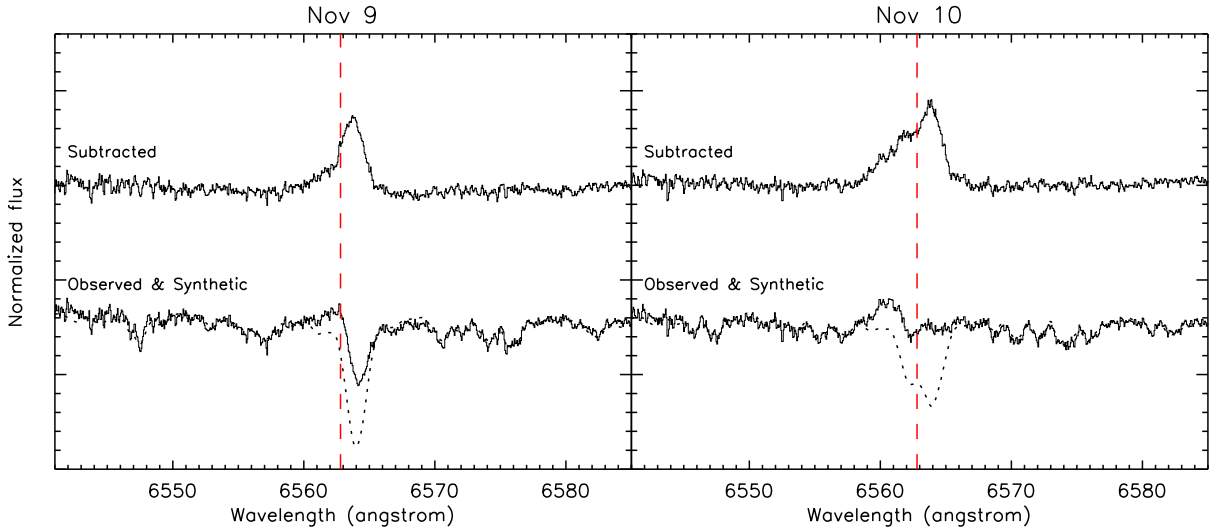} 
\caption{H$\alpha$ subtracted spectrum of TYC 8380-1953-1 for observations performed in 
Nov 9 and 10. Dashed lines are the synthetic spectra obtained with \textsc{jstarmod}. Continuous 
lines are used for the observed and the subtracted spectra. The vertical dashed line in both panels 
marks the laboratory wavelength for the H$\alpha$ line ($V_\mathrm{r} = 0$ km\,s$^{-1}$).}
\label{ha_sub}
\end{figure*}


\begin{figure*}
\includegraphics[width=17cm]{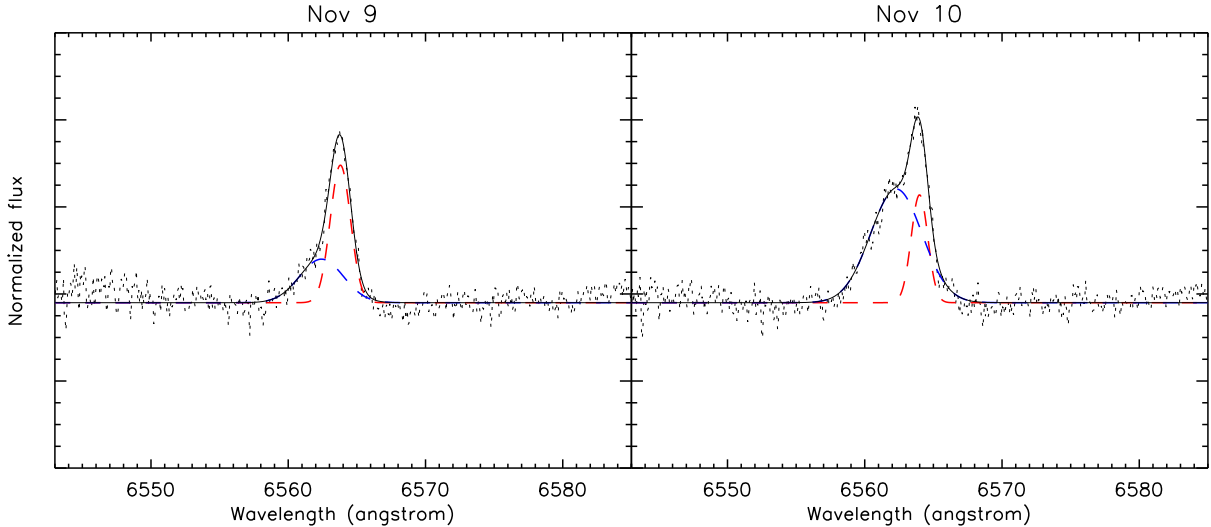} 
\caption{{ Modeling of the chromospheric H$\alpha$ emission of \tyc\ by two 
Gaussians.} In each panel, the subtracted
spectrum is plotted as a dotted-line. The continuous line is the fitted function
and the dashed lines are the two Gaussians corresponding to the two stars of 
the binary system.}
\label{ha_gauss}
\end{figure*}

\begin{figure*}
\includegraphics[width=17cm]{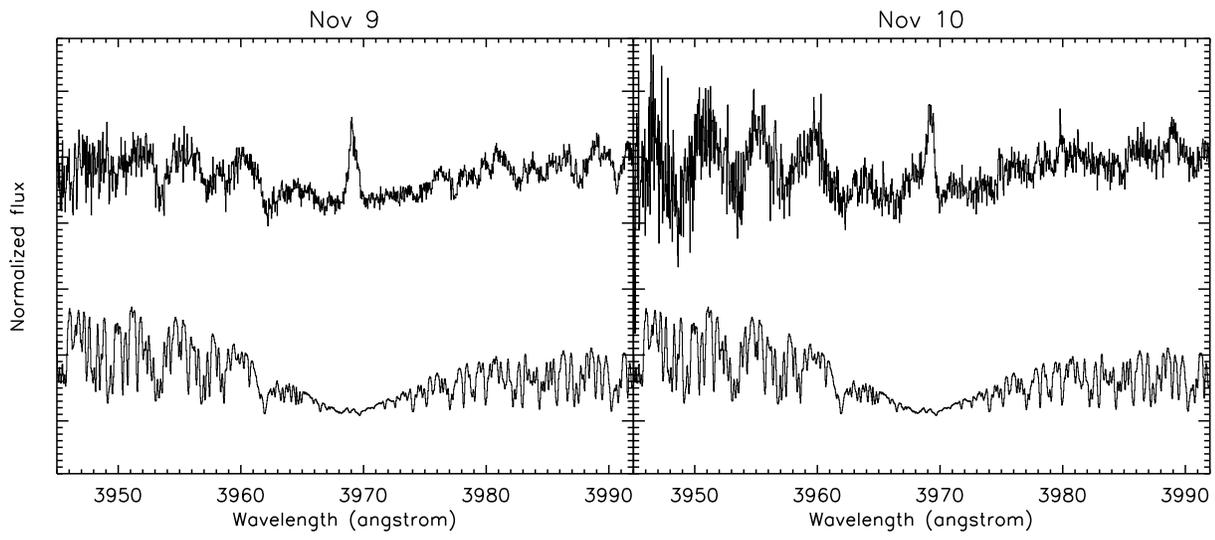} 
\caption{Normalized spectrum in the region of Ca~\textsc{ii} H for TYC 8380-1953-1
(up) and $\beta$ Cet (bottom) during Nov 9 and 10.}
\label{cah}
\end{figure*}

\begin{figure*}
\includegraphics[width=17cm]{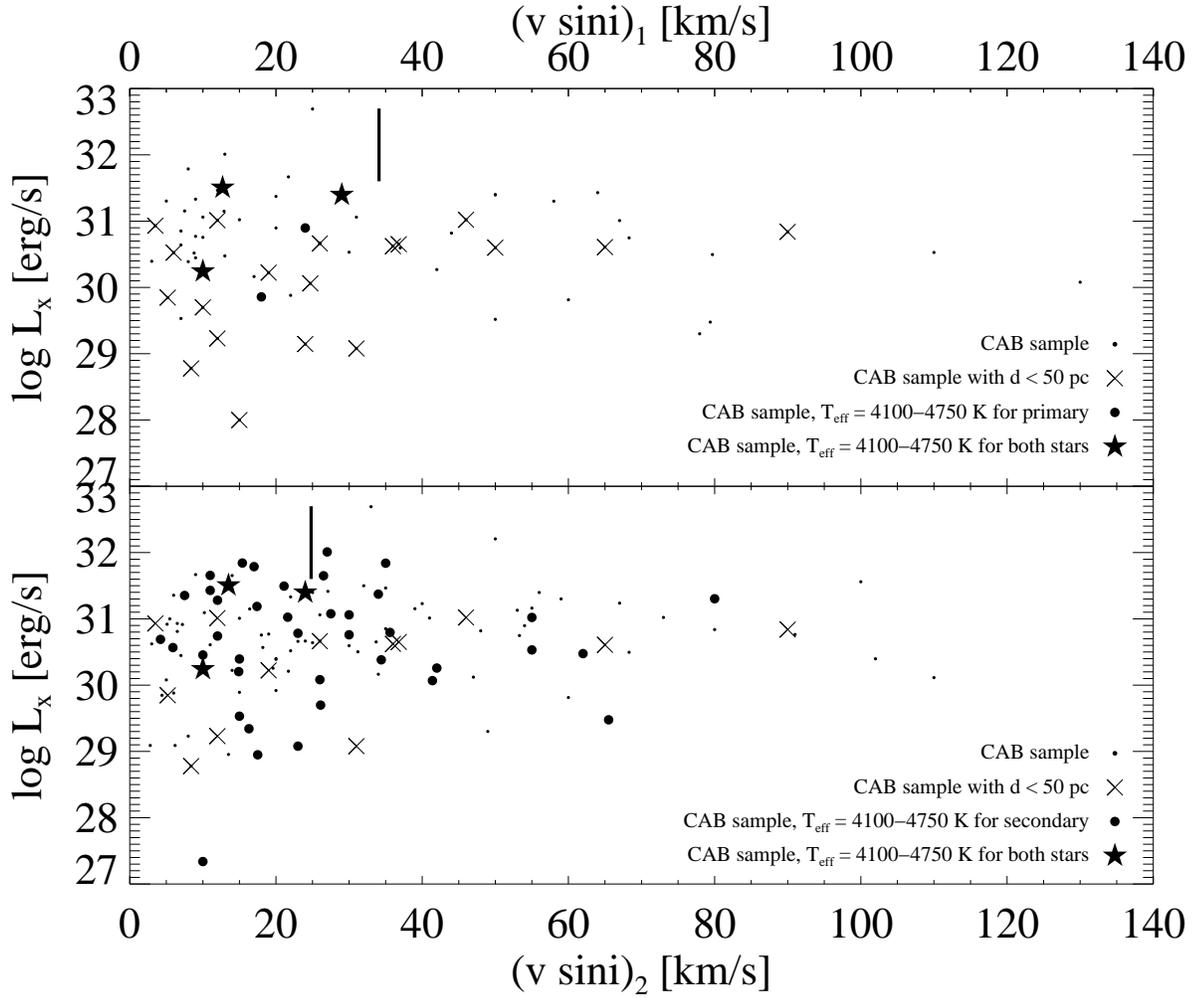} 
\caption{X-ray luminosity versus projected rotational velocity ($v \sin i$) for the primary and 
secondary components of binary systems in \citet{eke08}, top and bottom panels respectively.
\tyc\ is shown as vertical line for the range of $L_\mathrm{X}$ from the pointed
\textit{XMM-Newton} observation for d=600-2000 pc. 
}
\label{lx_vsini}
\end{figure*}



\end{document}